\documentclass[preprint,showpacs,preprintnumbers,amsmath,amssymb,nofootinbib,showkeys,aps]{revtex4}
\usepackage{graphicx}
\usepackage{dcolumn}
\usepackage{bm}
\usepackage{latexsym}
\usepackage{epsfig}
\usepackage{color}
\usepackage{hyperref}
\usepackage{bm}

\newcommand{\be}{\begin{equation}}
\newcommand{\ee}{\end{equation}}
\newcommand{\beqq}{\setlength\arraycolsep{2pt}\begin{eqnarray}}
\newcommand{\eeqq}{\vspace{0cm} \end{eqnarray}}
\newcommand{\bea}{\begin{eqnarray}}
\newcommand{\eea}{\end{eqnarray}}

\begin{document}

\title{Bayesian analysis of CCDM Models}

\author{J. F. Jesus} \email{jfjesus@itapeva.unesp.br}
\affiliation{Universidade Estadual Paulista (Unesp), C\^{a}mpus Experimental de Itapeva \\
Rua Geraldo Alckmin 519, 18409-010, Vila N. Sra. de F\'atima, Itapeva, SP, Brazil}

\author{R. Valentim}\email{valentim.rodolfo@unifesp.br}
\affiliation{Departamento de F\'{\i}sica, Instituto de Ci\^encias Ambientais, Qu\'{\i}micas e Farmac\^euticas - ICAQF, Universidade Federal de S\~ao Paulo (UNIFESP)
Unidade Jos\'e Alencar, Rua S\~ao Nicolau No. 210,
09913-030 -- Diadema, SP, Brazil}

\author{F. Andrade-Oliveira}\email{felipe.oliveira@port.ac.uk}
\affiliation{Institute of Cosmology and Gravitation -- University of Portsmouth\\
 Burnaby Road, PO1 3FX\\
Portsmouth, United Kingdom}


\begin{abstract}
Creation of Cold Dark Matter (CCDM), in the context of Einstein Field Equations, leads to negative creation pressure, which can be used to explain the accelerated expansion of the Universe. In this work we tested six different spatially flat models for matter creation using statistical tools, at light of SN Ia data: Akaike Information Criterion (AIC), Bayesian Information Criterion (BIC) and Bayesian Evidence (BE). These approaches allow to compare models considering goodness of fit and number of free parameters, penalizing excess of complexity. We find that JO model is slightly favoured over LJO/$\Lambda$CDM model, however, neither of these, nor $\Gamma=3\alpha H_0$ model can be discarded from the current analysis. Three other scenarios are discarded either from poor fitting, either from excess of free parameters.
\end{abstract}

%
\maketitle


\section{\label{introduction} Introduction}
Since evidence for cosmic expansion acceleration has been found \cite{riess1997,perlmutter,union21}, a large number of possible explanations for this unexpected behaviour have been proposed. 

The most accepted explanation, the $\Lambda$CDM model, is successful at explaining many observational data, e. g., the Cosmic Microwave Background Radiation \cite{planck}, Baryon Acoustic Oscillations \cite{bao} and $H(z)$ data \cite{hz}. However, it is plagued with the Cosmological Constant Problem \cite{weinberg89}, Cosmic Coincidence Problem \cite{Weinberg00} and some small scale problems \cite{Perivo08,Popolo16}.

In order to test other possibilities and trying to overcome the above difficulties, many alternatives to $\Lambda$CDM have been proposed. Among them, some models involve some kind of dark energy, with negative pressure, in order to convey acceleration inside Einstein Field Equations (EFE) and assuming the Cosmological Principle (CP) \cite{CopelandEtAl06}. Other proposals involve gravitation theory modifications \cite{SotiriouFaraoni10} or breaking of CP \cite{Celerier00}.

On the other hand, when it was first considered the possibility of quantum particle creation from evolving gravitational potentials \cite{prigogine89,CLW}, it was found that this particle creation led to the possibility of acceleration through the negative creation pressure, which rises from EFE with creation and assuming the CP validity.

As it was clarified on ref. \cite{CLW}, we may consider, as a simplifying hypothesis, that the particle creation should be `adiabatic'. As particle creation is essentially a non-adiabatic process, the total entropy increases, `adiabatic' here means that the entropy per particle number is conserved in this process.

Then, many so called Creation of Cold Dark Matter (CCDM) models were proposed in this scenario \cite{FreazaEtAl02,LSS08,ljo10,LimaBasilakos11,LimaEtAl12,JO16}, many of which were proposed phenomenologically through dimensional arguments about the particle creation rate. Recently, it was proposed a CCDM model which was equivalent to $\Lambda$CDM (the concordance model) with respect to the background equations, the so called LJO model \cite{ljo10}. Evolution of density perturbations were calculated to this model, in a Neo-Newtonian framework, and it was shown that this model can be distinguished from $\Lambda$CDM in the linear order, but it can be compatible with observations if some amount of entropy perturbations are considered \cite{jobl}. Later, it was shown that, even with no entropy perturbations, if one separates the obtained CDM density, through a part that clusters, and a non-clustered part, this model is equivalent to $\Lambda$CDM even at higher orders of density perturbation theory \cite{RamosEtAl14,SantosEtAl14}.

Meanwhile, more fundamental treatments of CCDM models were developed, through a particle creation rate deduced from a quantum particle creation in a curved spacetime \cite{jesuspereira}, and through the development of a kinetic theory of particle creation \cite{kinetic}.

The best way to compare the mentioned scenarios and to determine which mechanism has been driving the late stage of cosmic accelerated expansion, is using Bayesian criteria to differentiate among them in light of current observational data \cite{BayesCosm}.

In this work we consider the SNe Ia observations as it is the most straightforward evidence of cosmic acceleration and it counts to date with a great amount of data, enabling the best model selection in what concerns the background evolution equations.

Through the use of statistical tools namely, Bayesian Information Criterion (BIC), Akaike Information Criterion (AIC) and Bayesian Evidence (BE) we ranked some  models of interest, including those inside the CCDM scenario and the $\Lambda$CDM model. Our conclusions are drawn considering one of the largest Supernovae Ia data sample to date, the Union 2.1 \cite{union21}.

In Section II, we discuss the dynamics of the universe with negative pressure due to matter creation. In Section III, we discuss the model selection methods used here. In Section IV, we find the observational constraints from SNs Ia over selected CCDM models. In Section V, we apply the model selection methods to distinguish among CCDM models and compare with other results obtained in literature. Finally, we summarize the main results in conclusion.


\section{\label{model} Creation of Cold Dark Matter (CCDM) Models}

We will start by considering the homogeneous and isotropic FRW line element (with $c=1$):

\begin{equation}
\label{ds2}
  ds^2 = dt^2 - a^{2}(t) \left(\frac{dr^2}{1-k r^2} + r^2 d\theta^2+
      r^2{\rm sin}^{2}\theta d \phi^2\right),
\end{equation}
\noindent where $k$ can assume values $-1$, $+1$ or $0$.

In this background, the Einstein Field Equations are given by 

\begin{equation}
\label{fried1}
    8\pi G (\rho_{rad} + \rho_{b}  + \rho_{dm}) = 3 \frac{\dot{a}^2}{a^2} + 3 \frac{k}{a^2},
\end{equation}

\noindent and

\begin{equation}
\label{fried2}
   8\pi G (p_{rad} + p_{c}) =  -2 \frac{\ddot{a}}{a} - \frac{\dot{a}^2}{a^2} -
	\frac{k}{a^2}.
\end{equation}
where $\rho_{rad}$, $\rho_{b}$ and $\rho_{dm}$ are  the density parameters of radiation, baryons and dark matter,  $p_{rad}=\rho_{rad}/3$ is the radiation pressure and $p_c$ is the creation pressure.

The solutions of the EFE above are obtained considering an Energy-Momentum Tensor  (EMT) with the form \cite{prigogine89,kinetic}:

\begin{equation}\label{emtcorrected}
T^{\mu \nu} =T_{eq}^{\mu \nu}+\Delta T^{\mu \nu},
\end{equation}
where $T_{eq}^{\mu \nu}$ characterizes thermodynamic equilibrium in the fluid and the creation  of matter and entropy in universe are incorporated to the EFE through the correction term $\Delta T^{\mu \nu} = -p_c(g^{\mu\nu}-u^{\mu}u^{\nu})$ \cite{prigogine89,CLW,kinetic}.    

Therefore,  the complete EMT (\ref{emtcorrected}) in the presence of matter creation has  the explicit form:
\begin{equation}
T^{\mu \nu}=(\rho_{rad}+\rho_{b}+\rho_{dm}+p_{rad}+p_c)u^{\mu}u^{ \nu} -( p_{rad} + p_c) g^{\mu\nu}
\end{equation}

\noindent satisfying the conservation law $T^{\mu\nu}_{\,\,\,\,\,\,;\nu} = 0$.

Assuming solely the creation of dark matter component, the densities of radiation and baryon components  satisfy  their respective usual conservations laws, namely:

\begin{equation}\label{rad}
\dot{\rho}_{rad} +  4 \frac{\dot{a}}{a}\rho_{rad} = 0,
\end{equation}
and
\be\label{consbar}
\dot{\rho}_{b} + 3 \frac{\dot{a}}{a}{\rho}_{b} = 0.
\ee
where each overdot means one time derivative and we have used that $p_{rad}=\rho_{rad}/3$ and $p_{b}=0$.

On the other hand, when the creation process is considered we should take into account a matter creation source at level of Einstein Field Equations \cite{CLW}:

\begin{equation}
\label{ConsDM}
\frac{\dot{\rho}_{dm}}{{\rho}_{dm}} + 3 \frac{\dot{a}}{a} = \Gamma,
\end{equation}
where $\Gamma$ is the rate of dark matter creation in units of (time)${}^{-1}$.

As shown by \cite{CLW}, the creation rate of cold dark matter is associated to the creation pressure $p_c$ in Eq. (\ref{fried2}) through:

\begin{equation}
\label{pc}
  p_{c} = -\frac{\rho_{dm} \Gamma}{3H},
\end{equation}
where $H\equiv\dot{a}/a$ and we have considered an ``adiabatic" creation, i.e., the case when the entropy per particle is constant. The so called ``adiabatic'' regime is a simplifying hypothesis in which the only source of entropy increase in the universe is the matter creation \cite{prigogine89}. Mathematically, according to Calv\~ao, Lima \& Waga \cite{CLW}:

\begin{equation}
\dot{\sigma}=\frac{\Psi}{nT}\left(\beta-\frac{\rho+p}{n}\right)
\end{equation}
where $\sigma$ is the entropy per particle, $\Psi$ is the particle creation rate, $n$ is the particle density, $T$ is the temperature and $\beta$ is given from a phenomenological treatment of creation pressure:
\begin{equation}
p_c=-\frac{\beta\Psi}{\Theta}
\end{equation}
where $\Theta=3H$ is the bulk expansion rate. So, in case $\dot{\sigma}=0$, as we assume, we have $\beta=\frac{\rho+p}{n}$, then creation pressure is given by $p_c=-\frac{\rho+p}{\Theta}\frac{\Psi}{n}=-(\rho+p)\frac{\Gamma}{3H}$. On the other hand, if $\dot{\sigma}\neq0$, $\beta$ remains as an unknown parameter, which can not be constrained by thermodynamics alone, as the second law of thermodynamics demands only $\Psi\geq-\frac{n\dot{\sigma}}{\sigma}$.

As a consequence of Eq. (\ref{pc}), one can see that the dynamics of the universe is directly affected by the rate of creation of cold dark matter, $\Gamma$. In particular, in the case $\Gamma > 0$ (creation of particles) we have a negative pressure creation and in the case $\Gamma \to 0$  we recover the well known dynamics when the universe is lately dominated by pressureless matter (baryons plus dark matter). 

Since we are considering only the late phase of the dynamics of the universe, we can neglect the radiation terms from now on. Thus, by combining Eqs. (\ref{fried1}) and (\ref{fried2}), we have
\begin{equation}
 \frac{\ddot{a}}{a}=-\frac{4\pi G}{3}(\rho_{b} + \rho_{dm} +3p_c)
\end{equation}
Replacing $p_c$ from Eq. (\ref{pc}), we may write
\begin{equation}
 \frac{\ddot{a}}{a}=-\frac{4\pi G}{3}\left[\rho_{b} + \rho_{dm}\left(1-\frac{\Gamma}{H}\right)\right]
\end{equation}
Using that $\frac{\ddot{a}}{a}=\dot{H}+H^2$ and changing variables from time to redshift, we find
\begin{equation}
\frac{dH}{dz}=\frac{H}{1+z}+\frac{H_0^2\Omega_b(1+z)^2\Gamma}{2H^2}+\frac{H^2-H_0^2\Omega_k(1+z)^2}{2H(1+z)}\left(1-\frac{\Gamma}{H}\right)
\end{equation}
where we have used the solution of (\ref{consbar}) to baryon density, $\rho_b=\rho_{b0}(1+z)^3$, $\Omega_b=\frac{\rho_{b0}}{\rho_{c0}}$ is the present baryon density parameter, and $\Omega_k=-\frac{k}{H_0^2}$ is the present curvature density parameter. Changing to dimensionless variable $E(z)\equiv\frac{H(z)}{H_0}$, we find
\begin{equation}
\frac{dE}{dz}=\frac{E}{1+z}+\frac{\Omega_b(1+z)^2}{2E^2}\frac{\Gamma}{H_0}+\frac{E^2-\Omega_k(1+z)^2}{2E(1+z)}\left(1-\frac{\Gamma}{H_0E}\right)
\label{odeE}
\end{equation}
If the Universe is spatially flat, it can be further simplified:
\begin{equation}
\frac{dE}{dz}=\frac{3E}{2(1+z)}\left(1-\frac{\Gamma}{3H_0E}\right)+\frac{\Omega_b(1+z)^2}{2E^2}\frac{\Gamma}{H_0}
\label{odeEFlat}
\end{equation}
If we define the dimensionless quantity $\Delta\equiv\frac{\Gamma}{3H_0}$, it can be written:
\begin{equation}
\frac{dE}{dz}=\frac{3}{2}\left[\frac{E-\Delta}{1+z}+\frac{\Omega_b(1+z)^2}{E^2}\Delta\right]
\label{dEdz}
\end{equation}

So, if the CCDM model is defined with an expression $\Gamma=\Gamma(H)$, we can find a dependence $\Delta(E)$, replace it at Eq. (\ref{dEdz}) and solve it for $E(z)$.

\subsection{Models}
As we may consider $\Gamma(H)$ as a natural dependence for the creation rate, most of the CCDM models studied here follow this dependence. In fact, most of the models studied here can be written in a form $\Delta=\beta E+\alpha E^{-n}$, which corresponds to $\Gamma=3\beta H+3\alpha H_0\left(\frac{H_0}{H}\right)^n$. Another model we are interested in analyzing is the so called LJO model, with a dependence $\Gamma=3\alpha\frac{\rho_{c0}}{\rho_{dm}}H$, which has been shown to have the same background dynamics as the $\Lambda$CDM model. So, in this model, the cosmological constant is exactly mimicked by the creation of particles. The models we have analyzed are shown on Table \ref{tab1}.

\begin{table}[h!]
\renewcommand{\tabcolsep}{1.5pc} 
\renewcommand{\arraystretch}{1.2} 
	\centering
		\begin{tabular}{cccccc}
\hline\hline
Model &  Creation rate & Reference & Priors/Fixed Parameters\\
\hline

$M_0$	& $\Gamma=\frac{3\alpha H_0^2}{H}$ 									& JO \cite{JO16}																& $\alpha\in[0,1]$, $\beta=0$, $n=1$\\
$M_1$ & $\Gamma=3\alpha\frac{\rho_{c0}}{\rho_{dm}}H$ 			& LJO \cite{ljo10}
 	& $\alpha\in[0,1]$\\
$M_2$	& $\Gamma=3\alpha H_0$ 															&	\cite{GraefEtAl14}									& $\alpha\in[0,1]$, $\beta=0$, $n=0$\\
$M_3$	& $\Gamma=3\beta H$ 																&		--							& $\alpha=0$, $\beta\in[0,1]$\\
$M_4$	& $\Gamma=3\alpha H_0\left(\frac{H_0}{H}\right)^n$ 	&			--					& $\alpha\in[0,1]$, $\beta=0$, $n\in[-10,10]$\\
$M_5$	& $\Gamma=3\alpha\frac{H_0^2}{H}+3\beta H$ 					&	\cite{GraefEtAl14}									& $\alpha\in[-2,3]$, $\beta\in[-2,2]$, $n=1$\\
\hline
\hline
		\end{tabular}\\
\caption{Model parameters and priors.}
\label{tab1}
\end{table}

In all models analysed here we have taken into account the contribution of baryons. The baryon density was assumed to be a fixed parameter, given by Planck as $\Omega_b=0.049$. For simplicity, we choose to work with a spatially flat Universe, as indicated from CMB and preferred by inflation. So, $\Omega_k\equiv0$ in our analysis. Using the general expression for the creation rate, namely $\Gamma=3\beta H+3\alpha H_0\left(\frac{H_0}{H}\right)^n$,  Eq. (\ref{dEdz}) reads:
\begin{equation}
\frac{dE}{dz}=\frac{3}{2}\left[\frac{(1-\beta)E-\alpha E^{-n}}{1+z}+\Omega_b(1+z)^2\left(\frac{\alpha}{E^{n+2}}+\frac{\beta}{E}\right)\right]
\label{dEdz:mod}
\end{equation}
This equation covers all models studied here, except LJO ($M_1$). However, even neglecting spatial curvature, these models in general have no analytical expression for $H(z)$, due to baryons inclusion. One exception is LJO model, which recovers the $\Lambda$CDM dependence \cite{ljo10}:
\begin{equation}
 E(z)=\frac{H(z)}{H_0}=\left[\alpha+(1-\alpha)(1+z)^3\right]^{1/2}
\end{equation}
Another case in which $E(z)$ can be analytically obtained, even with the presence of baryons, is model $M_3$, where the creation rate is proportional to expansion rate. In this case:
\begin{equation}
 E(z)=\left[(1-\Omega_b)(1+z)^{3-3\beta}+\Omega_b(1+z)^3\right]^{1/2}
\end{equation}

If baryons parameter density $\Omega_b$ could be neglected, Eq. (\ref{dEdz:mod}) would yield:
\begin{equation}
 E(z)=\left[\frac{\alpha+(1-\alpha-\beta)(1+z)^{\frac{3}{2}(n+1)(1-\beta)}}{1-\beta}\right]^\frac{1}{n+1}
\end{equation}
However, baryon density brings an important contribution and can not be neglected. So, one has to resort to numerical methods or semi-analytical, approximated methods. Throughout our analysis, we solve Eq. (\ref{dEdz:mod}) numerically.

\section{\label{selection} Model Selection Methods}
In this section, we summarize the model selection methods used in this work. The likelihood function is the main part of analysis. It is built for each situation and there is not a general recipe for it \cite{ValentimEtAl11}. Here, we assume $N$ pairs of measurements $(x_i,y_i)$ for which we aim to find the best relation between $x$ and $y$. As a maximum likelihood estimator, we can use the $\chi^2$ expression given by \cite{SzydlowskiEtAl15}:
\begin{equation}
\chi^2 = \sum^{N}_{i=1}{\frac{\left[f(x_{i}, \theta)-y_{i}\right]^2}{\sigma_{i}^2}} = -2\ln{\mathcal{L}}.
\end{equation}
where $f(x_{i},\theta)$ represents the model with parameters $\theta$ and $\mathcal{L}$ is the likelihood function. The best fit values for the set of free parameters minimizes $\chi^2$ on the likelihood maximum. It is important to note that the likelihood function indicates the most likely values for the relevant parameters of the statistical model. 

\subsection{Ockham's razor}

Ockham's razor was proposed by William of Ockham (1285-1349) who was an English Franciscan friar and scholastic philosopher and theologian. The principle establishes: ``If there are two models with some parameters number, the most simple model will be better than more complex model''. This principle has two main reasons: aesthetic and empirical. First, if the model is more simple, it is more aesthetically beautiful because the most simple models would describe the Nature excluding null hypothesis. Second, the Nature is optimized i.e., it is economic and makes everything with parsimony. Ockham's razor is the main prerequisite for the construction of statistical models being used both as Frequentist and Bayesian analysis.

\begin{table}[t]
   \centering   
   \setlength{\arrayrulewidth}{2\arrayrulewidth}  
   \setlength{\belowcaptionskip}{10pt}  
   \begin{tabular}{ccccc} 
      \hline\hline
      $\Delta\mbox{AIC}$ & Support\\
      \hline
      $\Delta\mbox{AIC}\leq1$ &Not worth more than a bare mention\\
      $4\leq\Delta\mbox{AIC}\leq7$ & Significant/Weak\\
      $0\leq\Delta\mbox{AIC}\leq2$ & Strong to very strong/Significant\\
      $\Delta\mbox{AIC}>10$ & Decisive/Strong\\
     
      \hline \hline
   \end{tabular}
   \caption{Akaike Information Criterion}
   \label{tab:aic}
\end{table}

\subsection{Akaike Information Criterion}

The Akaike Information Criterion (AIC) is the relative measure of the quality of model with number of parameters and for a set of data. This criterion can estimate the quality of each model, relative to each of the other models. AIC is a kind of model selection and it emerges of Information Theory when there are not true models\footnote{This is not like saying that one model is true, but if one model is better than another through the goodness of fit \cite{Akaike74}.}, but it is only possible to obtain an approximation by models. Akaike \cite{Akaike74} created this criterion when he had studied approximate methods in models on Kullback-Leibler (KL) information. Akaike found an approximation to the KL quantity, which he called the Akaike Information Criterion (AIC), given by
\begin{equation}
\mbox{AIC} = -2\ln{\mathcal{L}_{max}} + 2p.
\end{equation}
where $\mathcal{L}_{max}$ is the maximum of the likelihood and $p$ is the number of model free parameters. This criterion can be used to compare different models. For two models with $\mbox{AIC}_1$ and $\mbox{AIC}_2$, we may define the difference:
\begin{equation}
\Delta\mbox{AIC} = \mbox{AIC}_2 - \mbox{AIC}_1.
\end{equation}

Table \ref{tab:aic} shows how to interpret the outcome of AIC. For two models, the likelihood of one model against the other is ``quantified'' by the difference $\Delta$AIC. This approach is suitable for understanding independently the goodness of fit. 

\subsection{Bayesian Information Criterion}

Bayesian Information Criterion (BIC) \cite{ValentimEtAl11,ValentimEtAl11b,SzydlowskiEtAl15} emerges from Bayes' Theorem, it is an approximation of Bayesian Evidence, and differently from AIC, this approach penalizes models with different number of free parameters. BIC takes into account Ockham's razor when it favours simple models against more complex models. BIC can be found as \cite{Schwarz78,Linddle04}:
\begin{equation}
\mbox{BIC} = -2\ln{\mathcal{L}_{max}} + p\ln{N}.
\end{equation}
where $N$ is the number of data, $\mathcal{L}_{max}$ is the maximum of likelihood and $p$ is the number of free parameters. The best model from a set under consideration is the one which minimizes the BIC quantity. The relative BIC is calculated by:
\begin{equation}
\Delta\mbox{BIC} = -(\mbox{BIC}_i-\mbox{BIC}_j).
\end{equation}

The interpretation of outcome is made in Table \ref{tab:bic}.


\begin{table}[t]
   \centering   
   \setlength{\arrayrulewidth}{2\arrayrulewidth}  
   \setlength{\belowcaptionskip}{10pt}  
   \begin{tabular}{ccccc} 
      \hline\hline
      $\Delta\mbox{BIC}$  & Support \\
      \hline
      $\Delta\mbox{BIC} \leq1$ &  No worth more than a bare mention \\
       $1\leq\Delta\mbox{BIC} \leq3$ &   Significant/Weak\\
      $3\leq\Delta\mbox{BIC} \leq5$ &  Strong to very strong/Significant\\
      $\Delta\mbox{BIC} >5$ &  Decisive/Strong \\
      \hline \hline
   \end{tabular}
   \caption{Bayesian Information Criterion}
   \label{tab:bic}
\end{table}

\subsection{Bayesian Evidence}
Bayesian Evidence (BE) emerges from Bayes' Theorem and it is product of two probability distributions: likelihood and prior distribution. The posterior probability function is defined by \cite{Kass95}:
\begin{equation}
P(M_i|D) = \frac{P(D|\theta,M_i)P(\theta_i|M_i)}{P(D)},
\end{equation}
where $P(M_i)$ is a prior probability for the model $M_i$. $D$ and $\theta_i$ denotes the data and free parameters, respectively. The $P(D)$ term is just a normalization. It is defined by:
\begin{equation}
P(D) = \int{P(D|\theta,M_i)P(\theta_i|M_i)d\theta}.
\end{equation}
$P(D)$ is calculated over all parameter space. The essential feature of Bayesian framework is the marginalization over all parameters, also called the Bayesian Evidence:
\begin{equation}
P(D|M_i) = \int{P(D|\theta,M_i)P(\theta_i|M_i)d\theta} = E_i.
\end{equation}
BE conveys the principle of Ockham's razor and allows to compare different models, through the Bayes factor \cite{Kass95,ElgMult06,GuimaraesEtAl09}:
\begin{equation}
B_{ij} =\frac{P(D|M_j)}{P(D|M_i)}.
\label{Bij}
\end{equation}

Note that in this definition we follow the convention of Ref. \cite{GuimaraesEtAl09}, in such a way that if $E(M_j)>E(M_i)$, then $\ln B_{ij}$ is positive. The interpretation of BE through the Bayes factor is shown on Table \ref{tabBE}.

\begin{table}[t]
   \centering   
   \setlength{\arrayrulewidth}{2\arrayrulewidth}  
   \setlength{\belowcaptionskip}{10pt}  
      \begin{tabular}{ccccc} 
      \hline\hline
      $\ln{B_{ij}}$ & Support\\
      \hline

      $\ln{B_{ij}}\leq$1 & Not worth more than a bare mention\\
      
       $1\leq\ln{B_{ij}}\leq2.5$ & Significant/Weak \\
      
       $2.5\leq\ln{B_{ij}}\leq5$  &  Strong to very strong/Significant \\
      
      $5<\ln{B_{ij}}$ & Decisive/Strong \\
      \hline \hline
   \end{tabular}
   \caption{Bayesian Evidence}
   \label{tabBE}
\end{table}

\section{\label{observational} Observational Constraints}
 In this section, we obtain constraints to the free parameters of the models listed on Table \ref{tab1}. In order to do this, we used the 580 Supernovae Type Ia (SN Ia) dataset of Union 2.1 \cite{union21}.

\subsection{Supernovae Type Ia  Bounds}

The parameters dependent distance modulus for a supernova at the redshift $z$ can be computed through the expression
\be
\label{mu}
\mu(z|\mathbf{s}) = m - M = 5\log d_L + 25,
\ee
where $m$ and $M$ are respectively the apparent and absolute magnitudes,  $\mathbf{s}\equiv (H_0, \alpha, \beta,n)$ is the set of the free parameters of the model and $d_L$ is the luminosity distance in unit of Megaparsecs.

Since in the general case $H(z)$ has not an analytic expression, we must define $d_L$ through a differential equation. The luminosity distance $d_L$ can be written in terms of a dimensionless comoving distance $D$ by:
\begin{equation}
d_L=(1+z)\frac{c}{H_0}D
\end{equation}

The comoving distance can be related to $H(z)$, neglecting spatial curvature, by the following relation \cite{clarkson}:
\be
\label{clark}
D'(z)=\frac{1}{E(z)},
\ee
where the prime denotes derivation with respect to redshift $z$. So, this equation, together with Eq. (\ref{dEdz:mod}) can be seen as a system of differential equations over the variables $(E(z),D(z))$. The initial conditions are, naturally, $(E(z=0)=1,D(z=0)=0)$.

We solve it numerically, then, in order to constrain the free parameters of the models, we considered the Union 2.1 SN Ia dataset from Suzuki {\it et al.} \cite{union21}. The best-fit  set of parameters $\mathbf{s}$ was estimated from a $\chi^2$ statistics with 

\be
\chi^2_{SN}=\sum^{N}_{i=1}\frac{\left[ \mu^i(z|\mathbf{s})-\mu^i_o(z)\right]^2}{\sigma_i^2}
\ee
where $\mu^i(z|\mathbf{s})$ is given by (\ref{mu}), $\mu^i_o(z)$ is the corrected distance modulus for a given SNe Ia at $z_i$ being $\sigma_i$ its corresponding individual uncertainty and $N=580$ for the Union 2.1 data compilation.

\begin{figure}[t]
\centerline{
\epsfig{figure=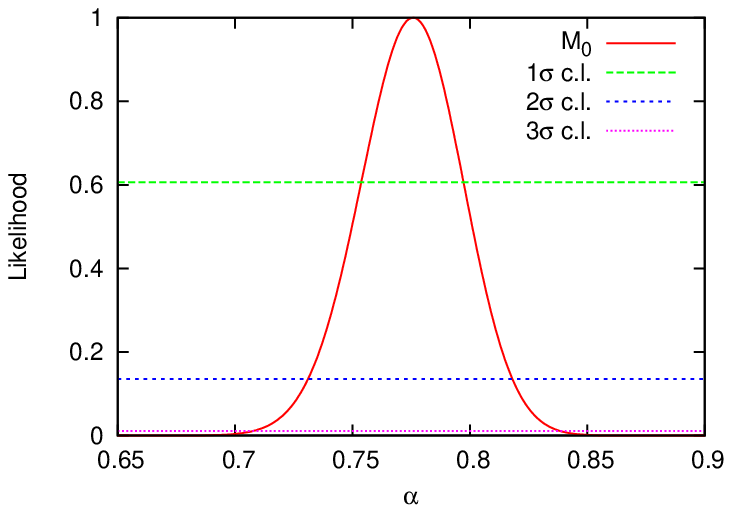,width=0.5\linewidth}
\epsfig{figure=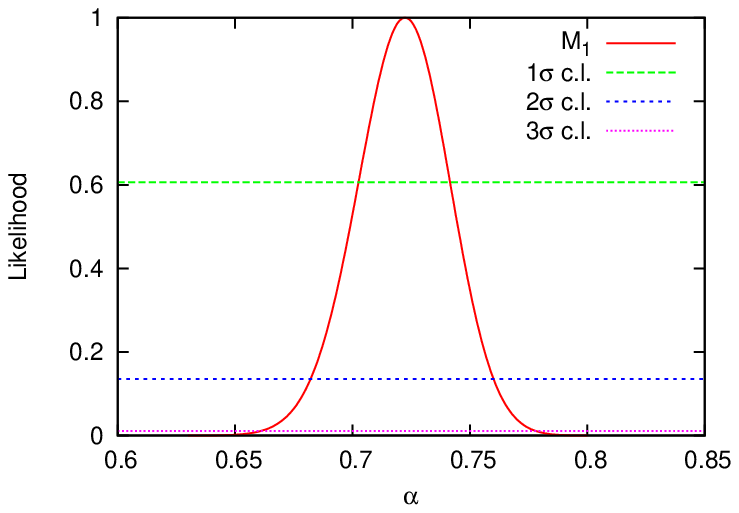,width=0.5\linewidth}}
\centerline{
\epsfig{figure=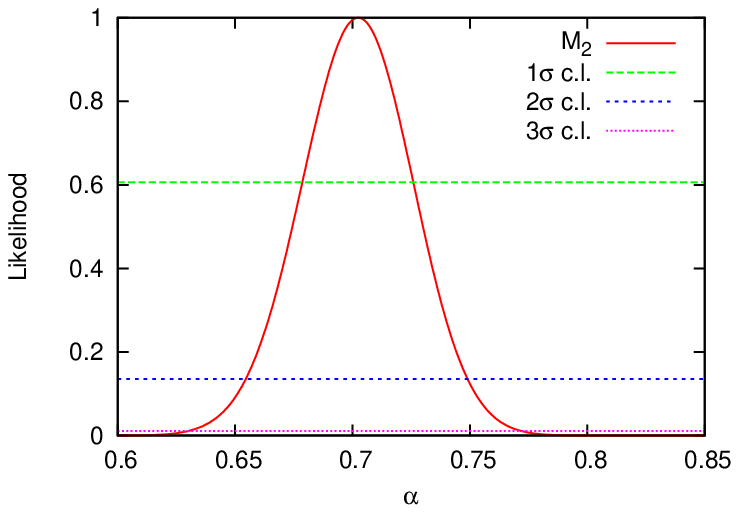,width=0.5\linewidth}
\epsfig{figure=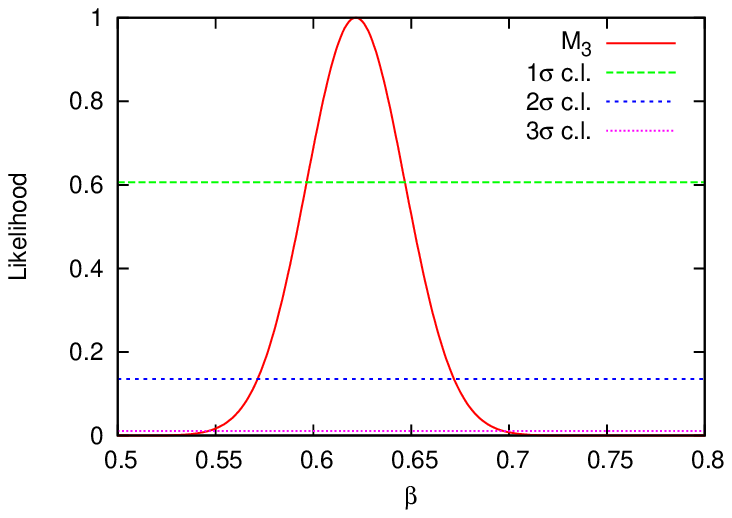,width=0.5\linewidth}
}
\centerline{
\epsfig{figure=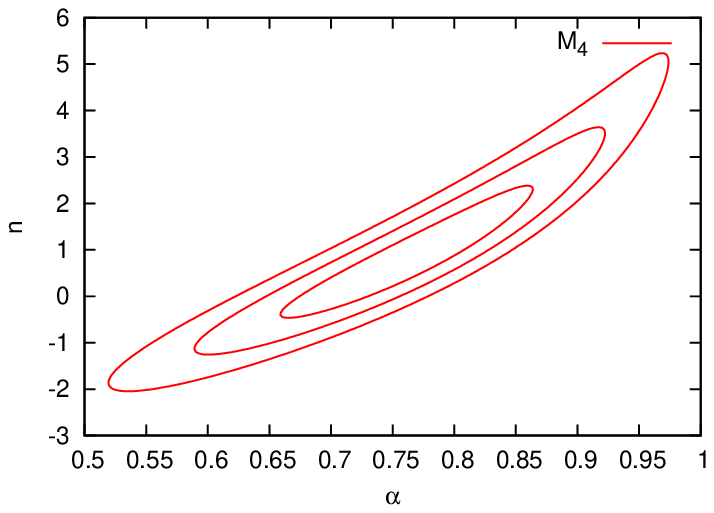,width=0.5\linewidth}
\epsfig{figure=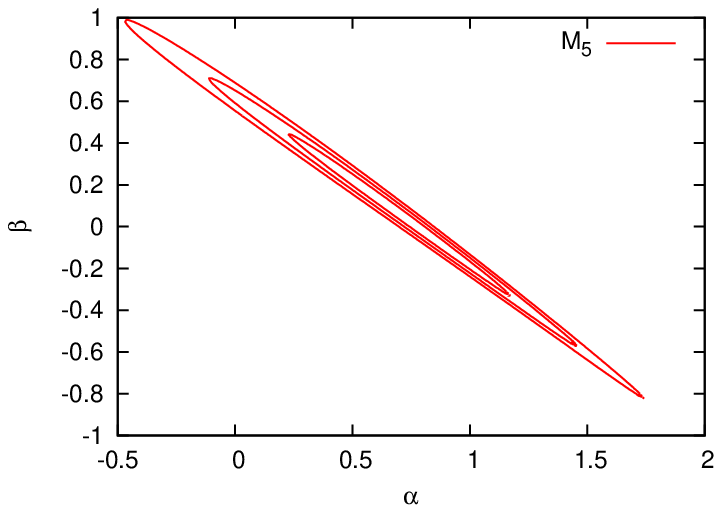,width=0.5\linewidth}}
\caption{The results of our statistical analysis, with constraints from SNs Union 2.1 data. {\bf Panels (a)-(d)} Likelihoods for the parameters on each indicated model, $M_0$-$M_3$. {\bf Panels (e)-(f)} Contours for 68.3\%, 95.4\% and 99.7\% confidence intervals for each indicated model, $M_4$ and $M_5$. Explanation of each model is on text and Table \ref{tab1}.}
\label{contours}
\end{figure}

As usual on SN Ia analyses, we rewrite the distance modulus:
\be
\mu = 5\log(D_L) + M_*
\ee
where $D_L=(1+z)D$ is dimensionless luminosity distance and $M_*\equiv25+5\log\frac{c}{H_0}$ comprises all the dependence over $H_0$. Then, we marginalize the likelihood over $M_*$:
\be
\tilde{\mathcal{L}}(\alpha,\beta,n)=\mathcal{N}\int_{-\infty}^{+\infty}\exp\left[-\frac{1}{2}\chi^2(M_*,\alpha,\beta,n)\right]dM_*
\ee
where $\mathcal{N}$ is a normalization constant. The corresponding $\tilde{\chi}^2=-2\ln\left(\frac{\tilde{\mathcal{L}}}{\mathcal{N}}\right)$ is given by:
\begin{equation}
 \tilde{\chi}^2=C-\frac{B^2}{A}
\end{equation}
where
\begin{equation}
A=\sum_{i=1}^{N}\frac{1}{\sigma_i^2},\quad B=\sum_{i=1}^{N}\frac{5\log[D_L(z_i)]-\mu_{o,i}}{\sigma_i^2},\quad C=\sum_{i=1}^{N}\left\{\frac{5\log[D_L(z_i)]-\mu_{o,i}}{\sigma_i}\right\}^2. 
\end{equation}
The result of this analysis can be seen on Fig. \ref{contours}.

As one may see on Fig. \ref{contours}, the models are well constrained by SN Ia Union 2.1 data. For panels {\bf (a)}-{\bf (d)} we can see the likelihood $\mathcal{L}$ for parameters $\alpha$ and $\beta$ of models $M_0$ to $M_3$, respectively. In panel {\bf (e)}, we see the likelihood contours for model $M_4$, with free parameters $\alpha$ and $n$, corresponding to 68.3\%, 95.4\% and 99.7\% c.l. The same corresponding contours we may see on Panel {\bf (f)} for model $M_5$, with free parameters $\alpha$ and $\beta$. The detailed results for each model can be seen on Table \ref{tab2}.

\begin{table}[ht]
\centering
\renewcommand{\arraystretch}{1.2}
\setlength{\tabcolsep}{5pt}
  \begin{tabular}{  ccccccccccccc}
    \hline\hline
    Model 						& $\alpha$ & $\beta$ & $n$ & $\chi^2_\nu$\\ \hline
$M_0:$ $\Gamma=\frac{3\alpha H_0^2}{H}$ 		& $0.776_{-0.022}^{+0.021}$	& 0 	& 1 	&0.97107	\\
$M_1:$ $\Gamma=3\alpha\frac{\rho_{c0}}{\rho_{dm}}H$ 	&  $0.722_{-0.020}^{+0.019}$	& -- 	& --	&0.97103	\\
$M_2:$ $\Gamma=3\alpha H_0$ 				& $0.702_{-0.024}^{+0.023}$	& 0	& 0	&0.97259	\\
$M_3:$ $\Gamma=3\beta H$ 				& 0	& $0.622\pm0.025$	& --	&0.97916	\\
$M_4:$ $\Gamma=3\alpha H_0\left(\frac{H_0}{H}\right)^n$ & $0.766_{-0.11}^{+0.098}$	& 0	& $0.8_{-1.3}^{+1.5}$	&0.97270	\\
$M_5:$ $\Gamma=3\alpha\frac{H_0^2}{H}+3\beta H$ 	& $0.74_{-0.34}^{+0.28}$	& $0.03_{-0.23}^{+0.27}$	& 1	&0.97270	\\    
\hline
\hline
  \end{tabular}
\caption{Results of the analysis for the different models. Limits on the parameters correspond to 68.3\% c.l. as explained on text.}
\label{tab2}
\end{table}

\newpage

\section{Model selection of matter creation models}
Next, we have calculated AIC, BIC and Bayesian Evidence for all models studied here. AIC and BIC are relatively easy to calculate, as they are directly obtained from $\chi^2_{min}$. As explained above, the Akaike Information Criterion (AIC) emerged from information theory like an approximation to Kullback-Leibler (KL) information \cite{Akaike74,SzydlowskiEtAl15}. It is given by
\begin{equation}
\mathrm{AIC}=-2\ln\mathcal{L}_{max}+2p
\end{equation}
In our case, $\mathcal{L}=\mathcal{N}\exp\left(-\frac{\chi^2}{2}\right)$, where $\mathcal{N}$ is a normalization constant, thus
\begin{equation}
\mathrm{AIC}=\chi^2_{min}-2\ln\mathcal{N}+2p
\end{equation}
The normalization constant cancels out when we calculate the difference $\Delta$AIC between two different models:
\begin{equation}
\Delta\mathrm{AIC}=\mathrm{AIC}_j-\mathrm{AIC}_i=\chi^2_{min,j}-\chi^2_{min,i}+2(p_j-p_i)
\end{equation}

The $\Delta$AIC values for the six models studied here are shown on third column of Table \ref{tab3}. In the second column, we have the values of reduced chi-square, $\chi^2_\nu=\chi^2_{min}/\nu$. The values of $\chi^2_\nu$ varies little for all the models studied here, favouring slightly the model $M_1$, the so called LJO model, which gives the same background dynamics as $\Lambda$CDM. $\Delta$AIC goes in the same direction and indicates an slightly preference for LJO. The values of $\Delta$AIC in this column are relative to $M_0$, the CCDM model where $\Gamma\propto\frac{1}{H}$. Let us call it JO, for short. The models that are excluded by the Akaike criterion are $M_3$, $M_4$ and $M_5$, but mainly $M_3$, due to its high $\chi^2$ value. Because AIC penalizes too weakly the number of free parameters, it favours $M_4$ and $M_5$ over $M_3$ because they provide a lower $\chi^2_{min}$, although $M_4$ and $M_5$ have more free parameters than $M_3$. A Bayesian criterion, one which drastically penalizes the excess of free parameters is thus necessary.

\begin{table}[ht]
\centering
\renewcommand{\arraystretch}{1.2}
\setlength{\tabcolsep}{5pt}
  \begin{tabular}{ccccccccccc}
    \hline\hline
    Model 										&$\chi^2_{min}$& $\chi^2_\nu$ 			& $\Delta$AIC 	& $\Delta$BIC	& $V_P$ & $\ln B_{0i}$\\ \hline
$M_0:$ $\Gamma=
\frac{3\alpha H_0^2}{H}$ 			&	562.251 & 0.97107						&  0						&  	0					& 1			& 0						\\
$M_1:$ $\Gamma=3\alpha
\frac{\rho_{c0}}{\rho_{dm}}H$ &	562.227 & 0.97103						& -0.024 				&	-0.024			& 1			& 0.043				\\
$M_2:$ $\Gamma=3\alpha H_0$ 	&	563.131 & 0.97259						& 0.880					& 	0.880			& 1			& 0.155				\\
$M_3:$ $\Gamma=3\beta H$ 			&	566.936 & 0.97916						& 4.685					& 	4.685			& 1			& 0.955				\\
$M_4:$ $\Gamma=3\alpha H_0
\left(\frac{H_0}{H}\right)^n$ &	562.221 &	0.97270						& 1.969					& 	6.332			& 20		& 0.921			\\
$M_5:$ $\Gamma=3\alpha
\frac{H_0^2}{H}+3\beta H$ 		&	562.220 &	0.97270 					& 1.969					& 	6.332			& 20		& 1.463				\\    
\hline
\hline
  \end{tabular}
\caption{Results of the model selection analysis for the different models.}
\label{tab3}
\end{table}

As the Bayesian Evidence is, in general, hard to evaluate, as it is given by multidimensional integrals over the parameters, Schwarz has proposed an approximation to it, known as BIC. The Bayesian Information Criterion (BIC) is seemingly easy to calculate, as it is given by
\begin{equation}
\mathrm{BIC}=-2\ln\mathcal{L}_{max}+p\ln N
\end{equation}
As an approximation to Bayesian Evidence, BIC drastically penalizes the excess of free parameters for a big sample, such as the SN Ia sample. In our case, BIC is given by
\begin{equation}
\mathrm{BIC}=\chi^2_{min}-2\ln\mathcal{N}+p\ln N
\end{equation}
and the normalization constant $\mathcal{N}$ is cancelled out on $\Delta$BIC:
\begin{equation}
\Delta\mathrm{BIC}=\mathrm{BIC}_j-\mathrm{BIC}_i=\chi^2_{min,j}-\chi^2_{min,i}+(p_j-p_i)\ln N
\label{deltabic}
\end{equation}

The values of $\Delta$BIC for CCDM are shown on fourth column of Table \ref{tab3}. As one may see, BIC excludes model $M_3$ due to bad fitting ($\Delta\chi^2=4.685$ relatively to $M_0$) and it excludes $M_4$ and $M_5$ due to excess of parameters. In fact, although the $\chi^2_{min}$ is slightly lower for these models, this small advantage is quite penalized due to the term $\ln580=6.363$ in BIC equation (\ref{deltabic}). We may say that the great effort of adding one free parameter in $M_4$ or $M_5$ can not be justified by the small improvement of fitting obtained.

However, as we know, BIC is only an approximation of the Bayesian evidence. Bayesian evidence is the most reliable tool to perform model selection, in the Bayesian Statistics point of view. Let us calculate the Bayesian evidences then.

As discussed on \cite{GuimaraesEtAl09}, the Bayesian evidence can be written as
\begin{equation}
E=\int\mathcal{L}(p)\pi(p)dp
\end{equation}
where $\pi(p)$ is the prior probability distribution for the parameters. Assuming flat priors, we may write:
\begin{equation}
E=\frac{1}{V_\pi}\int_{V_\pi}\mathcal{L}(p)dp
\end{equation}
where $V_\pi$ is the volume in the parameter space spanned by the prior. As one may see, the Bayesian evidence can be quite dependent over the prior choice, even if it is flat. However, as discussed on \cite{ElgMult06}, this dependence is weaker if one chooses large prior intervals. In fact, if the prior volume is large enough to encompass all the region that the likelihood is non-negligible, one can ensure that the logarithmic Bayesian evidence grows linearly with the logarithmic prior volume, as expected. Based on this, we choose conservative priors for the model parameters, ensuring that the 3$\sigma$ likelihood constraints are quite inside of the prior volume region. In some cases, we had to limit the priors with physical considerations like to avoid big bangless models, we must have $\alpha\in[0,1]$. However, these limits were always inside the 3$\sigma$ likelihood constraints, as one may see on Fig. \ref{contours}.

We have compared the CCDM models by using the Bayes factor (\ref{Bij}). As mentioned before, we use a convention where $\ln B_{ij}$ is positive in case that $E_j>E_i$. The results of this analysis for the models studied here is in the seventh column of Table \ref{tab3}, where we show the values of $B_{0i}$, the Bayes factors relatively to model $M_0$ (JO).

As one may see on Table \ref{tab3}, while models $M_3$ and $M_4$ were barely acceptable in comparison with model $M_0$, the only model that can be discarded by this analysis is model $M_5$.

In order to compare our results with the current literature, \cite{GraefEtAl14} have obtained $\Delta$AIC for three models analyzed here, $M_1$, $M_2$ and $M_5$ (CCDM1, CCDM2 and CCDM3 in their analysis, respectively). They have calculated $\Delta$AIC relatively to $M_1$. Comparing with their analysis, our result for $M_5$ is quite similar. However, we did not find such a large $\Delta$AIC $=33.21$ as they have found. Our $\Delta$AIC for $M_5$, relatively to $M_1$, is $\Delta$AIC $=1.993$.

\section{\label{conclusion} Conclusion}
We have compared 6 spatially flat CCDM models, one of which is indistinguishable from the $\Lambda$CDM model. The JO model is slightly favoured over $\Lambda$CDM in the Bayesian evidence, however, $\Lambda$CDM and $\Gamma=3\alpha H_0$ can not be discarded from this analysis. Models $M_3$ and $M_4$ can be moderately discarded and $M_5$ can certainly be discarded. At this point, it is important to mention that JO model is equivalent to the late phase of the model from Ref. \cite{LimaEtAl12}.

Further investigations may include spatial curvature, other background data and evolution of density perturbations.

\begin{acknowledgments}
The authors wish to thank A. C. C. Guimar\~aes for very helpful discussions. FAO is supported by CNPq-Brazil through a fellowship within the program Science without Borders and  R. V. has been supported by FAPESP (Process no. 2013/26258-4). 
\end{acknowledgments}

\end{document}